\documentclass[prc,aps,twocolumn,floatfix,groupedaddress,nofootinbib,preprintnumbers,amsmath,amssymb
,amsfonts,superscriptaddress,widetable,href]{revtex4-1}
\usepackage{amsfonts}
\usepackage{amsmath}
\usepackage{amssymb}
\usepackage{graphicx,color}
\usepackage{rotating}
\usepackage{dcolumn}
\usepackage{bm,ulem,soul}
\usepackage{color}
\definecolor{navyblue}{rgb}{0.3,0.3,1}
\definecolor{purple}{rgb}{0.6,0,0.5}
\usepackage[colorlinks=true,pdfstartview=FitV,linkcolor=blue,citecolor=blue,urlcolor=navyblue]{
hyperref}

\begin{document}

\title{Analysis of critical parameters for nonrelativistic models in symmetric nuclear matter} 

\author{Mariana Dutra}
\address{Departamento de F\'isica, Instituto Tecnol\'ogico de Aeron\'autica, DCTA, 
12228-900, S\~ao Jos\'e dos Campos, SP, Brazil}

\author{Odilon Louren\c{c}o}
\email{odilon.ita@gmail.com}
\address{Departamento de F\'isica, Instituto Tecnol\'ogico de Aeron\'autica, DCTA, 
12228-900, S\~ao Jos\'e dos Campos, SP, Brazil}

\author{Xavier Vi\~nas}
\address{Departament de F\'isica Qu\`antica i Astrof\'isica and Institut de 
Ci\`encies del Cosmos (ICCUB), Facultat de F\'isica,Universitat de Barcelona, Mart\'i i Franqu\`es 
1, E-08028 Barcelona, Spain}

\author{C. Mondal}
\address{Departament de F\'isica Qu\`antica i Astrof\'isica and Institut de 
Ci\`encies del Cosmos (ICCUB), Facultat de F\'isica,Universitat de Barcelona, Mart\'i i Franqu\`es 
1, E-08028 Barcelona, Spain}
\address{LPC Caen, Universit\'e de Caen Normandie, F-14000, Caen, France.}

\date{\today}

\begin{abstract}
In this work we have analyzed several features of symmetric nuclear matter (SNM) at finite 
temperature described by different zero- and finite-range nonrelativistic families of models, 
namely, Skyrme, Gogny, 
Momentum-dependent interaction (MDI), Michigan three-range Yukawa (M3Y) and Simple Effective 
Interaction (SEI). We have calculated the critical parameters (CP) associated to the liquid-gas 
phase coexistence for nuclear matter from these 
parametrizations and show that they are in agreement with their experimental and theoretical values 
obtained in the literature. Our study also points out to a strong evidence of universality 
presented by the hadronic models, namely, model independence in the gaseous phase and 
distinguishability among different interactions in the liquid phase. We have performed a correlation 
study among different CP and SNM properties. Such studies involving different finite range 
interactions are scarce in literature. The analyzed models show an overall increasing trend of the 
critical temperature as a function of critical pressure.
\end{abstract}

\maketitle

\section{Introduction}
Hadronic models show very interesting features when they are used to describe warm nuclear matter 
at non-vanishing temperatures. Due to nucleon-nucleon interaction, thermodynamical liquid and gas 
phases coexist below a certain temperature, named critical temperature, exhibiting a van der Waals 
pattern. The different thermodynamical quantities such as the pressure ($P_c$) or the density 
($\rho_c$) at this junction point along with the temperature $T_c$ are denoted together as critical 
parameters. The analysis of such phase structure can lead to a deeper understanding of the nuclear 
interaction in different environments such as heavy-ion collisions~\cite{hic1,hic2} and finite 
nuclei~\cite{borderie,natowitz,sfienti}, for instance. In non-accreting neutron stars, finite 
temperature calculations might also play some crucial roles to determine the structure and 
composition of their crusts~\cite{Fantina19, Carreau19}. Furthermore, a suitable knowledge of the 
hadronic equations of state at $T>0$ is crucial to describe correctly different astrophysical 
phenomena such as the core-collapse supernovae or the neutron star mergers \cite{prakash,andre}. 
Correlation studies among the nuclear matter properties (at $T=0$) and critical parameters ($T>0$) 
also carry vital information regarding the nuclear equation of state and in turn the basic nature of 
the nucleon-nucleon interaction in medium \cite{PRC95-065212}. In other words, if strong enough 
correlations are established between $T_c$, $P_c$ and $\rho_c$ with the bulk parameters of hadronic 
models, any direct or indirect experimental constraints established in a particular set of these 
quantities might be useful to pin down the other ones.

Equations of state are obtained by using nuclear models of different degrees of 
sophistication. Explaining nuclear phenomena based on a theory starting from fundamental 
nucleon-nucleon interaction is yet to be achieved. The nonperturbative nature of the nuclear force 
makes it very difficult to be described starting from the strong interaction between 
quarks and gluons. Over the years, developing effective theories by optimizing few parameters 
{fitted to} certain experimental data has been a hallmark of the development 
of the nuclear theory. As 
an example, the ground state energy of a nucleus, which is defined as the negative of its binding 
energy $B(A,Z)$, was proposed long time ago by Weizs\"acker~\cite{drop} in a model (semi-empirical 
mass formula) that considers the nucleus as a droplet of incompressible matter with $B(A,Z)$ 
containing terms proportional to its volume, surface, etc \cite{benzaid}. Many 
sophisticated models have been constructed since then, successfully describing different features 
of 
finite nuclei and infinite nuclear matter. Some of them, not necessarily in the chronological order 
of appearing, are described in the following. 

In {\it Chiral effective field theory (EFT) models} (see Ref.~\cite{eft} and references therein), 
the 
most general Lagrangian density is proposed with the basic symmetries of quantum chromodynamics, 
{in particular} the chiral symmetry. The low-energy regime of this theory is obtained 
with the quarks confined into the colorless hadrons giving rise to the more suitable degrees of 
freedom 
for this energy scale~\cite{eft}. However, nuclear forces based on chiral EFT also pose some major 
challenges to be applied in nuclear structure \cite{machleidt16} and reactions {(see 
\cite{Whitehead20}
and references therein).}

In its effective finite range version, the {\it Relativistic mean-field (RMF) models} explicitly 
describe the attractive and repulsive nuclear interactions by including in the Lagrangian density 
the fermion field $\psi$ coupled to the scalar and vector mesons fields $\sigma$ and $\omega_\mu$, 
respectively. The structure of the model also generates scalar and vector potentials that largely 
cancel each other at a particular density, giving rise to a relativistic mechanism for the nuclear 
matter saturation. In its point-coupling version, the RMF models consider a zero range interaction 
between the nucleons, and $\psi$ is the {only} field in this case. For the finite, zero range, 
and even improved versions of this model, see Refs. \cite{rev1,rev2,rev3}. The density dependent 
meson 
exchange (DDME) version of the RMF models also describe several ground state finite nuclear 
properties 
satisfactorily~\cite{vretner03}.

The nonrelativistic {\it Skyrme model} considers nucleons interacting each other through two and 
three-body point-like interactions~\cite{sky1,sky2,sky3,sky4,stoneskyrme}. Its two-body potential 
is written as a contact term times a low-momentum expansion function, taken up to quadratic order in 
the momenta. The three-body interaction is given by the product of two delta functions, {which} can be 
also interpreted as a two-body density dependent interaction. Different thermodynamical 
quantities in Skyrme models can be obtained in a relatively simple way, as they can be expressed as 
functions of the nuclear density and the proton fraction (zero temperature regime). Such ease in 
implementation has made Skyrme models so popular over the years. It has also been successfully 
applied to finite nuclei reproducing with very good accuracy the ground states energies, giant 
resonances and other physical properties 
\cite{Guo-Quiang91}.

The standard {\it Gogny models} of the D1 family consist of two finite-range terms of Gaussian 
type, which include all the possible spin-isospin exchange operators with different weights, plus a 
zero-range density-dependent contribution~\cite{gogny1}. The main property of Gogny interaction is 
that it can describe simultaneously the mean field and the pairing field. The Gogny 
interactions correctly describe many features of finite nuclei, in particular their pairing and 
deformation properties, along the whole periodic table~\cite{gogny2,gogny3,gogny4}. Although Gogny 
forces of D1 family do not describe properly the properties of neutron stars~\cite{gogny6,gogny7}, 
recent progress has been made in this direction~\cite{gogny8,gognyic}. 

The {\it Momentum-dependent interaction (MDI)} was primarily designed to be used in heavy ion 
collisions~\cite{mdi1}. Similar to the Gogny forces, the MDI can be written as a single 
finite-range 
term of Yukawa type, along with two zero-range contributions. Although the first versions of the 
MDIs 
were adequate to describe collisions of symmetric nuclei, more recent versions of this force are 
able to describe collisions of neutron-rich nuclei at intermediate 
energies~\cite{mdi2,mdi3}. The MDI is constructed in such a way that it is possible to obtain a 
family of forces with the same properties of symmetric matter but with variation in the isovector 
sector of the force \cite{mdi2,mdi3}, which makes these interactions very appealing to be used in 
the calculations of neutron star properties~\cite{mdi4}. 

The so-called {\it M3Y models} (Michigan three-range Yukawa)~\cite{m3y1} were derived from a bare 
nucleon-nucleon interaction (Paris, Reid) by fitting the microscopic $G$-matrix to the sum of three 
Yukawa form-factor of different ranges acting on the different spin-isospin states. It should be 
pointed out that the original M3Y force was unable to reproduce the saturation and spin-orbit 
splitting at mean field level. To solve this problem, zero-range terms were added and some 
strengths were modified~\cite{m3y2}. The tensor force, which is important for describing the shell 
structure in finite nuclei, has been included in the different M3Y 
parametrizations~\cite{m3y2,m3y3,m3y4,m3y5}. In order to describe open-shell nuclei, pairing 
correlations have been taken into account using the M3Y force in the particle-particle channel 
together with a cutoff in momentum space~\cite{m3y3}. It is also important to mention that the 
interactions of the M3Y type have been applied successfully to describe different nuclear 
reactions~\cite{m3y6}.

The {\it Simple Effective Interaction (SEI)} was constructed in 1998 by Behera and 
collaborators~\cite{behera98} aimed to describe nuclear and neutron star matter at zero and finite 
temperatures. The SEI consists of a single finite-range term with a form-factor of Gauss or Yukawa 
type, a pure contact term and a zero-range density-dependent contribution, which contains an 
additional parameter to avoid the supraluminous behavior at any temperature~\cite{behera98}. At 
variance with other effective interactions like Skyrme, Gogny or M3Y type, nine out of the eleven 
parameters of SEI are fitted to empirical constraints and microscopic results obtained with 
realistic interactions in nuclear matter. In this way SEI predicts the correct behavior of the 
momentum dependence of the mean field as extracted from heavy-ion collisions at intermediate 
energies. SEI also predicts trends of Dirac-Brueckner-Hartree-Fock and variational calculations in 
nuclear and neutron matter. One of the remaining two parameters is fixed from the microscopic 
spin-up spin-down splitting of the effective mass in polarized neutron matter~\cite{behera15}. The 
last parameter, together with the strength of the spin-orbit contribution, are determined from 
Hartree-Fock calculations in finite nuclei~\cite{behera13,behera16}. It is worthwhile to point out 
that, in spite of the fact that almost all the parameters of SEI are determined in nuclear matter, 
its finite nuclei description has a quality similar to that found using successful effective 
interactions like Skyrme, Gogny or M3Y.

In a previous investigation, we have used RMF models to calculate different characteristics of 
nuclear matter at finite temperature~\cite{PRC95-065212}. In the present work, we intend to 
complement that study with calculations performed for different nonrelativistic models. To this 
end, we study the nonrelativistic Skyrme, Gogny, MDI, M3Y and SEI models at finite temperature 
regime in order to compute different critical parameters and compare them with experimental and 
theoretical results. We also investigate the connection between these quantities 
with some bulk parameters, namely, incompressibility and nucleon effective mass, both calculated 
for 
symmetric nuclear matter at zero temperature. In Sec.~\ref{nrel} we outline the 
main theoretical quantities regarding the nonrelativistic models studied in this work (expressions 
at finite temperature). The outcomes of the finite temperature calculations are shown in 
Sec.~\ref{results} and in Sec.~\ref{summary}, a short summary and our concluding remarks are 
presented.

\section{Nonrelativistic models at finite temperature}
\label{nrel}

\subsection{Skyrme model}
An advantage of the Skyrme model is that its point-like nucleon-nucleon interaction implies a 
Hamiltonian as a function only of the nuclear density $\rho$ for symmetric systems. In the mean-field 
approach, the single particle state of the nucleon in a uniform medium is written in terms of a 
plane-waves \cite{latt}. As a consequence, it is straightforward to construct, at zero temperature, the energy 
density of the system and therefore to derive all the other thermodynamical quantities needed to 
describe nuclear matter, see, for instance, Ref.~\cite{PRC85-035201} for such calculations. 

At finite temperature regime the Heaviside step function ($\theta(k_F-k)$ with $k_F$ being the 
Fermi momentum in units of fm$^{-1}$) present in all the momentum integrals at zero temperature, is replaced 
by the Fermi-Dirac function (momentum distribution), depending on momentum $k$, temperature $T$ and an effective 
chemical potential $\mu$, which is given by 
\begin{eqnarray}
n_{\mbox{\tiny sky}}(k)  = \frac{1}{e^{[\varepsilon^*(k)-\mu]/T}+1},
\label{fd1}
\end{eqnarray}
where $\varepsilon^*(k)=\hbar^2k^2/2M^*$ is the single-particle energy with $M^*$ being the 
effective 
mass. As a consequence, for warm 
nuclear matter the nuclear density 
becomes~\cite{PRC63-044605,kuo}
\begin{eqnarray}
\rho = \frac{\gamma}{(2\pi)^3}\int d{\bf k}\,n_{\mbox{\tiny sky}}(k),
\label{rho}
\end{eqnarray}
where $\gamma$ is the degeneracy factor ($\gamma$=4 for symmetric nuclear matter).
For the Skyrme model, the nucleon effective mass in the single-particle energy $\varepsilon^*$ is 
defined from the energy density as its variation with respect to the kinetic energy density. It is 
given by
\begin{eqnarray}
M^* &=& M\left[1 +
\frac{1}{8}\frac{M}{\hbar^2}\rho\left(3t_1 + 5t_2 + 4t_2x_2\right) \right]^{-1},
\end{eqnarray}
in which $M = 939$~MeV is the free nucleon mass. Notice that in Eq.~(\ref{rho}) the momentum 
distribution 
depends on $M^*$ instead of $M$. For the numerical calculations, the van der Waals-like isotherms 
are obtained for a fixed temperature and run over the density. For a particular $\rho$, we 
invert Eq.~(\ref{rho}) in order to find the value of the chemical 
potential $\mu$. Then, for each $\rho$ we can compute the momentum distribution, which 
enters in all the other thermodynamical quantities, since the corresponding value of 
$\mu$ 
is determined. As we are interested in the critical parameters of the model, obtained through the 
following conditions 
\begin{eqnarray}
P_c=P(\rho_c , T_c),\quad 
\frac{\partial P}{\partial\rho}\bigg|_{\rho_c , T_c}=0,\quad
\frac{\partial^2 P}{\partial\rho^2}\bigg|_{\rho_c , T_c}=0,\quad
\label{conditions}
\end{eqnarray}
it is only needed to construct the pressure of the system, since it is the most relevant 
thermodynamical quantity for this purpose. For the Skyrme model it reads
\begin{align}
P_{\mbox{\tiny sky}}(\rho,T) &= \frac{3t_0}{8}\rho^2 + 
\frac{1}{16}\sum_{i=1}^{3}t_{3i}(\sigma_i+1)\rho^{\sigma_i+2} 
\nonumber \\
&+
\frac{\gamma\hbar^2}{6\pi^2M^*}
\left(1-\frac{3}{2}\frac{\rho}{M^*}\frac{dM^*}{d\rho}\right)
\int_0^\infty dk\,k^4 n_{\mbox{\tiny sky}}(k).
\label{presskyrme}
\end{align}

For the symmetric system, a particular parametrization of the Skyrme model is defined by a specific 
set of the following free parameters: $x_2$, $t_0$~[MeV.fm$^3$], $t_1$~[MeV.fm$^5$], 
$t_2$~[MeV.fm$^5$], $t_{3i}$~[MeV.fm$^{3(\sigma_i+1)}$], and~$\sigma_i$. Here, we mainly focus on 
the Consistent Skyrme parametrizations (CSkP) selected in Ref~\cite{PRC85-035201}. In that work, 
the authors select $16$ Skyrme parametrizations that satisfies the 11 constraints coming from 
nuclear matter, pure neutron matter, analysis of symmetry energy and its derivatives. They are: 
GSkI~\cite{agrawal2006}, GSkII~\cite{agrawal2006}, KDE0v1~\cite{agrawal2005}, LNS~\cite{cao2006}, 
MSL0~\cite{chen2010}, NRAPR~ \cite{steiner2005}, Ska25s20~\cite{private2}, Ska35s20~\cite{private2}, 
SKRA~\cite{rashdan2000}, Skxs20~\cite{brown2007}, SQMC650~\cite{guichon2006}, 
SQMC700~\cite{guichon2006}, SkT1~\cite{tondeur1984}, SkT2~\cite{tondeur1984}, 
SkT3~\cite{tondeur1984} and SV-sym32~\cite{klupfel2009}. Among these parametrizations, 
only two are ``nonstandard'', namely, GSkI and GSkII. The term nonstandard refers here to those 
parametrizations for which $i$ is not equal to 1 in Eq.~(\ref{presskyrme}). In particular, GSkI and 
GSkII were shown to fit consistently the masses of some spherical nuclei, namely, $^{16}\rm O$, 
$^{24}\rm O$, $^{14}\rm Ca$, $^{48}\rm Ca$, $^{48}\rm  Ni$, $^{56}\rm  Ni$, $^{68}\rm 
Ni$, $^{78}\rm Ni$, $^{88}\rm Sr$, $^{90}\rm Zr$, $^{100}\rm Sn$, $^{132}\rm  Sn$, and $^{208}\rm  
Pb$. The CSkP was also shown to be consistent~\cite{epja} with the constraints extracted from the 
LIGO and Virgo Collaboration analysis, related to the detection of gravitational waves coming from 
the neutron star merger GW170817 event~\cite{ligo17,ligo18,ligo19}. {For the sake of 
completeness, we also add to our analysis 4 more Skyrme parametrizations. Three of them are 
constrained by chiral effective field theory~\cite{lim17}, namely, \mbox{Sk$\chi414$}, 
\mbox{Sk$\chi450$} and \mbox{Sk$\chi500$}, and another one taken from Ref.~\cite{malik19}, 
\mbox{Sk$\Lambda 267$}. For the latter one, the dimensionless tidal deformability of the 
$1.4M_\odot$ neutron star is given by $\Lambda_{1.4}=267$, with the corresponding radius of 
$R_{1.4}=11.6$~km.}

\subsection{Finite-range interactions}

The finite-range (FR) interactions that we study in this work, namely Gogny, MDI, M3Y and SEI have 
a similar structure, which can be written as 
\begin{align}
&V(\mathbf{r}_1,\mathbf{r}_2) = \sum_{i=1}^N (W_i + B_i P_{\sigma} - H_i P_{\tau} - M_i 
P_{\sigma}P_{\tau})f(r,\mu_i)
\nonumber\\
	&+ t_0 (1+ x_0 P^\sigma) \rho^{\alpha_0} (\mathbf{R}) \delta (\mathbf{r})
	+ t_3 (1+ x_3 P^\sigma) \rho^{\alpha_3} (\mathbf{R}) \delta (\mathbf{r}),
\label{VGogny}     
\end{align}
where ${\bf r}={\bf r_1}-{\bf r_2}$ and ${\bf R}=({\bf r_1}+{\bf r_2})/2$ are the relative and the 
center of mass coordinates. $W_i$, $B_i$, $H_i$ and $M_i$ are the strengths of all the possible 
combinations of the spin ($P^{\sigma}$) and isospin exchange ($P^{\tau}$) operators, respectively.
 $\mu_i$ are the ranges of the $N$ form-factor 
(Gaussian for Gogny and Yukawian for MDI or M3Y {and can be both for SEI}) that describe the 
finite-range part of the force ($N$=1 for MDI {or SEI}, $N$=2 for Gogny and $N$=3 for M3Y). In 
Eq.(\ref{VGogny}) we have neglected the spin-orbit and tensor parts of the interaction owing to the 
fact that they do not contribute to the infinite nuclear matter.

In the case of warm {symmetric} nuclear matter described by  a finite-range interaction given in 
Eq.~(\ref{VGogny}), the single-particle energy is given by (see for instance Ref.~\cite{temp1})
\begin{align}
\varepsilon(k) &=
\frac{\hbar^2 k^2}{2M} + \frac{3}{8}t_0\alpha_0t_0 \rho^{\alpha_0-1} + 
 \frac{3}{8}\alpha_3t_3\rho^{\alpha_3-1}
\nonumber \\
&+ \sum_{i=1}^N g(0,\mu_i)\left[W_i + \frac{B_i}{2} - \frac{H_i}{2} - \frac{M_i}{4}\right]\rho
\nonumber \\
&+ \sum_{i=1}^N\left[M_i + \frac{H_i}{2} - \frac{B_i}{2} - \frac{W_i}{4}\right]\times
\nonumber\\
&\frac{\gamma}{(2\pi)^3}\int d{\bf k'}n(k')\tilde{g}(k,k',\mu_i),
\label{spe}
\end{align}
where $\tilde{g}(k,k',\mu_i)$ is the angular averaged Fourier transform of the finite-range form 
factor $f({\bf r - r'},\mu_i)$ \cite{temp1} (see Appendix for more details) and $\gamma$ the 
degeneracy factor introduced before. The momentum distribution $n(k)$ in Eq.~(\ref{spe}) is given by
\begin{eqnarray}
n(k) = \frac{1}{e^{[\varepsilon(k)-\mu]/T}+1}.
\label{fd2}
\end{eqnarray}
At difference with the case of zero-range forces in which the integral of the momentum distribution
$n(k)$ in Eq.~(\ref{rho}) determines the effective chemical potential, in the case of finite-range 
forces one needs to solve the coupled system of Eqs.~(\ref{spe}) and~(\ref{fd2}) with the 
constraint 
of Eq.~(\ref{rho}), which for a given density allows to obtain the chemical potential $\mu$.
Once the Fermi-Dirac occupation number $n(k)$ is determined by this procedure, one can 
easily determine the energy density as
\begin{eqnarray}
&&{\cal H} =
\frac{\gamma}{(2\pi)^3} \int d{\bf k}\frac{\hbar^2 k^2}{2M} + \frac{3}{8}t_0\rho^{\alpha_0} +
 \frac{3}{8}t_3\rho^{\alpha_3}
\nonumber \\
&&+ \frac{1}{2}\sum_{i=1}^N g(0,\mu_i)\bigg[W_i + \frac{B_i}{2} - \frac{H_i}{2} - 
\frac{M_i}{4}\bigg]\rho^2
\nonumber \\
&&+ \sum_{i=1}^N\bigg[M_i + \frac{H_i}{2} - \frac{B_i}{2} - \frac{W_i}{4}\bigg]\times
\nonumber \\
&&\frac{\gamma^2}{2}\int \frac{d{\bf k}}{(2\pi)^3}n(k)\int \frac{d{\bf 
k'}}{(2\pi)^3}n(k')\tilde{g}(k,k',\mu_i),
\label{ener}
\end{eqnarray} 
and the entropy density as
\begin{align}
{\cal S} &= \gamma  \int \frac{d{\bf k}}{(2\pi)^3}\lbrace n(k) \ln[n(k)] + [1 - 
n(k)]\ln[1-n(k)]\rbrace
\nonumber \\
&= \frac{1}{T}\int \frac{d{\bf k}}{(2\pi)^3}n(k)\bigg[ \varepsilon(k) + \frac{k}{3}\frac{d 
\varepsilon(k)}{dk}\bigg].
\label{entro}
\end{align}
Finally, the pressure at a given temperature $T$ is given by the standard thermodynamical relation
\begin{eqnarray}
P_{\mbox{\tiny FR}}(\rho,T) = \mu\rho- {\cal F} = \mu\rho - {\cal H} + {\cal S}T,
\label{press}
\end{eqnarray}
where ${\cal F}$ is the free energy density.

It is important to {mention here} that in the case of SEI, the second density-dependent term in 
Eq.~(\ref{ener}) is divided by a factor $(1 + b\rho)^{\alpha_3-2}$ and the contribution to the
corresponding single-particle energy (\ref{spe}) is also modified accordingly. We label the SEI 
parametrizations used in this work by $G$ or $Y$ to indicate if the form factor is of the Gauss or 
Yukawa type and by the value of the corresponding incompressibility modulus. More details about 
these parametrizations can be found in Refs.~\cite{behera15,routray16}.

\section{Analysis of the finite temperature calculations}
\label{results}

Before discussing the results in details, we make some general remarks about the 
nuclear matter properties of the models chosen for our study. The zero-range and finite-range 
mean-field models used in this study, in general, reproduce reasonably well binding energies and 
charge radii of finite  nuclei and predict nuclear matter properties usually within the window of the 
empirical values, namely, energy per nucleon $e_0=-15.8 \pm 0.5$ MeV, saturation density 
$\rho_0=0.16 \pm 0.01$ fm$^{-3}$, isoscalar effective mass ratio $m^*=M^*(\rho_0)/M=0.6-1.0$ and 
incompressibility modulus $K_0=240 \pm 30$~MeV (see for instance Ref.~\cite{malik19}). We emphasize 
here the importance of the saturation density~$\rho_0$, since it is directly related to the short 
range nature of the nuclear force. Because of this feature, protons and neutrons only interact with 
their near surrounding nucleons and this mechanism leads to approximately constant value of 
$\rho_0$. Regarding the Gogny interactions considered in this work, we see that there are some 
parametrization with incompressibility modulus outside the window of the empirical values 
(see Table \ref{tabcritical}). Among these the D1S interaction was fabricated to build up an accurate 
mass table \cite{hilaire08}. The rest of the 
parametrizations with high $K_0$ values were built up in Ref.~\cite{blaizot95} in order to study the 
correlation between the incompressibility modulus in nuclear matter and the energy of the monopole 
vibrations. The isoscalar effective mass ratio $m^*$ of the finite-range models considered in this 
work lie in the range of $0.6-0.7$, which reproduce the excitation energy of the isoscalar giant 
quadrupole resonance \cite{bohigas79}. This value of the isoscalar effective mass is in agreement 
with the value extracted from the optical model analysis of the nucleon-nucleus scattering 
\cite{li15}. Some of the Skyrme models which we have considered, predict an effective mass 
close to the bare mass. Models with an effective mass ratio equal or slightly larger than unity 
predict a single-particle level density close to the Fermi surface, which is in good agreement with 
the experiment without considering an additional particle-vibration coupling~\cite{dutta86}. 
However, these models with effective mass close to the bare mass are prone to predict maximum masses 
of neutron stars below the lower limit of the observed value of 2.01$\pm 0.04 M_{\odot}$~\cite{malik19}. 

Since the pressure as a function of $\rho$ and $T$ of zero and finite range interactions is 
determined, as shown in Eqs.~(\ref{presskyrme}) and (\ref{press}), it is now possible to analyze 
the critical parameters and the main features of the thermal symmetric nuclear matter for the 
different nonrelativistic models introduced  in the previous section. Nevertheless, before 
that, a comment regarding the phase transition in nuclear systems is needed 
at this point. Conjectures concerning the existence of a liquid-gas phase transition in strongly 
interacting matter have been corroborated through indirect evidences, since the critical point 
itself can not be directly observed in nuclear experiments. One of such evidences involves the 
distribution of the intermediate mass fragments produced, for instance, in the 
following reactions: $^{84}\rm Kr + ^{197}Au$~\cite{imf1}, $\rm Au + C$~\cite{imf2}, $\rm Au + 
Al$~\cite{imf2}, $\rm Au + Cu$~\cite{imf2}, $^{197}\rm Au+^{197}\rm Au$~\cite{imf3}, $p+\rm 
Xe$~\cite{imf4} and $p+\rm Kr$~\cite{imf4}. Another possible signature of the nuclear phase 
transition is identified from the analysis of the so called ``caloric curve'', or in other words,  
the dependence of temperature on the excitation energy per particle in finite nuclei. It was first 
predicted theoretically in Ref.~\cite{bondorf} and later discovered by the \mbox{ALADIN}
collaboration~\cite{aladin}, from a fragment distributions study produced in $\rm Au+Au$ collisions 
at incident energy of 600~MeV per nucleon. The plateau exhibited by this curve is characteristic 
of systems presenting phase transitions, thus supporting the existence of such thermodynamical 
phenomenology in nuclear systems~\cite{nato}.

We start by showing in Table~\ref{tabcritical} the critical parameters $P_c$, $\rho_c$ and $T_c$ 
along with $\rho_c/\rho_0$, the compressibility factor $Z_c=P_c/\rho_cT_c$, and some bulk 
parameters, namely, incompressibility $K_0$, isoscalar effective mass ratio 
$m^*$ (at $\rho=\rho_0$), and the saturation density itself~($\rho_0$).  
Concerning the ratio~$Z_c$, one can verify that all parametrizations present $Z_c$ smaller than the 
respective value related to the van der Waals model, namely,~$0.375$. This is a feature also 
observed for relativistic models, as pointed out, for instance, in Refs.~\cite{vdw5,PRC95-065212}. 
Notice that Table \ref{tabcritical} enlists, for the first time, to the best of our knowledge, the 
critical parameters $T_c$, $P_c$ and $\rho_c$ for almost all the non-relativistic finite range 
effective nucleon-nucleon interactions available in the literature.
\begin{table}[!htb]
\begin{ruledtabular}
\caption{Critical parameters $T_c$ (MeV), $\rho_c$ (fm$^{-3}$) and $P_c$ (MeV/fm$^3$), along 
with the quantities, namely, $\rho_c/\rho_0$, $Z_c=P_c/\rho_cT_c$, and the bulk 
parameters $K_0$ (MeV), $m^*=M^*(\rho_0)/M$, and $\rho_0$ (fm$^{-3}$) for different 
nonrelativistic parametrizations used in this work.}
\scriptsize
\centering
\begin{tabular}{l|c|c|c|c|c|c|c|c}
Model & $T_c$ & $\rho_c$ & $P_c$        &~~$\frac{\rho_c}{\rho_0}$~~ & ~~$Z_c$~~ & $K_0$ &~ $m^*$ 
& 
$\rho_0$\\
\hline
GSkI     & 15.09 & 0.052 & 0.223 & 0.328 & 0.284 & 230.21 & 0.776 & 0.159\\ 
GSkII    & 15.27 & 0.052 & 0.226 & 0.328 & 0.284 & 233.40 & 0.790 & 0.159\\ 
KDE0v1   & 14.86 & 0.054 & 0.225 & 0.330 & 0.279 & 227.54 & 0.744 & 0.165\\ 
LNS      & 14.93 & 0.057 & 0.235 & 0.328 & 0.275 & 210.78 & 0.826 & 0.175\\ 
MSL0     & 15.17 & 0.053 & 0.226 & 0.330 & 0.282 & 230.00 & 0.800 & 0.160\\ 
NRAPR    & 14.39 & 0.054 & 0.218 & 0.337 & 0.280 & 225.65 & 0.694 & 0.161\\ 
Ska25s20 & 16.27 & 0.053 & 0.239 & 0.329 & 0.278 & 220.75 & 0.980 & 0.161\\ 
Ska35s20 & 17.16 & 0.054 & 0.264 & 0.339 & 0.287 & 240.27 & 1.000 & 0.158\\ 
SKRA     & 14.36 & 0.052 & 0.208 & 0.329 & 0.276 & 216.98 & 0.748 & 0.159\\ 
SkT1     & 17.06 & 0.055 & 0.266 & 0.339 & 0.286 & 236.16 & 1.000 & 0.161\\ 
SkT2     & 17.04 & 0.055 & 0.265 & 0.339 & 0.286 & 235.73 & 1.000 & 0.161\\ 
SkT3     & 17.04 & 0.055 & 0.265 & 0.339 & 0.286 & 235.74 & 1.000 & 0.161\\ 
Skxs20   & 15.38 & 0.052 & 0.216 & 0.321 & 0.270 & 201.95 & 0.964 & 0.162\\
SQMC650  & 14.85 & 0.057 & 0.234 & 0.331 & 0.277 & 218.11 & 0.779 & 0.172\\  
SQMC700  & 14.73 & 0.057 & 0.233 & 0.332 & 0.278 & 222.20 & 0.755 & 0.171\\  
SV-sym32 & 16.03 & 0.053 & 0.242 & 0.332 & 0.285 & 233.81 & 0.900 & 0.159\\  
Sk$\chi$414	& 18.33 & 0.059 & 0.311	& 0.349 & 0.286 & 243.18 & 1.075 & 0.170\\ 
Sk$\chi$450	& 17.08 & 0.053 & 0.261	& 0.341 & 0.287 & 239.54 & 1.006 & 0.156\\ 
Sk$\chi$500	& 18.20 & 0.059 & 0.305	& 0.349 & 0.285 & 238.16 & 1.087 & 0.168\\ 
Sk$\Lambda$267	& 14.63 & 0.054 & 0.224	& 0.337 & 0.281 & 230.08 & 0.702 & 0.162\\ 
&  &  &  &  &  &  &  & \\  
D1S      & 15.89 & 0.060 & 0.281 & 0.368 & 0.295 & 202.88 & 0.697 & 0.163\\  
D1M      & 15.95 & 0.058 & 0.272 & 0.352 & 0.294 & 224.98 & 0.746 & 0.165\\  
D1N      & 15.76 & 0.056 & 0.261 & 0.348 & 0.296 & 225.65 & 0.747 & 0.161\\  
D250     & 17.16 & 0.061 & 0.332 & 0.386 & 0.318 & 249.54 & 0.702 & 0.158\\ 
D260     & 15.48 & 0.059 & 0.273 & 0.369 & 0.299 & 259.49 & 0.615 & 0.160\\ 
D280     & 15.21 & 0.058 & 0.263 & 0.379 & 0.298 & 285.19 & 0.575 & 0.153\\ 
D300     & 16.80 & 0.058 & 0.310 & 0.372 & 0.318 & 299.14 & 0.681 & 0.156\\  
&  &  &  &  &  &  &  & \\  
MDI      & 15.62 & 0.058 & 0.268 & 0.363 & 0.296 & 210.98 & 0.673 & 0.160\\  
 &  &  &  &  &  &  &  & \\  
M3Y-P1   & 15.78 & 0.062 & 0.294 & 0.367 & 0.301 & 225.70 & 0.641 & 0.169\\  
M3Y-P2   & 15.66 & 0.059 & 0.277 & 0.363 & 0.300 & 220.40 & 0.652 & 0.163\\  
M3Y-P3   & 16.12 & 0.060 & 0.290 & 0.369 & 0.300 & 245.80 & 0.658 & 0.163\\  
M3Y-P4   & 16.08 & 0.060 & 0.294 & 0.369 & 0.304 & 235.30 & 0.665 & 0.163\\  
M3Y-P5   & 15.78 & 0.060 & 0.289 & 0.369 & 0.305 & 235.60 & 0.629 & 0.163\\  
M3Y-P4'  & 15.91 & 0.060 & 0.290 & 0.369 & 0.304 & 230.40 & 0.653 & 0.163\\
M3Y-P5'  & 15.88 & 0.060 & 0.291 & 0.369 & 0.306 & 239.10 & 0.637 & 0.163\\ 
M3Y-P6   & 15.97 & 0.061 & 0.306 & 0.375 & 0.314 & 239.70 & 0.596 & 0.163\\  
M3Y-P7   & 16.33 & 0.062 & 0.326 & 0.381 & 0.322 & 254.70 & 0.589 & 0.163\\  
 &  &  &  &  &  &  &  & \\ 
SEIG263  & 16.30 & 0.056 & 0.278 & 0.361 & 0.305 & 262.52 & 0.712 & 0.155\\ 
SEIG245  & 15.79 & 0.055 & 0.260 & 0.350 & 0.300 & 245.62 & 0.711 & 0.157\\ 
SEIG227  & 15.23 & 0.055 & 0.242 & 0.344 & 0.289 & 227.64 & 0.710 & 0.160\\ 
SEIG207  & 14.55 & 0.054 & 0.221 & 0.333 & 0.281 & 207.69 & 0.709 & 0.162\\ 
SEIY282  & 17.35 & 0.061 & 0.340 & 0.379 & 0.321 & 282.30 & 0.686 & 0.161\\ 
SEIY254  & 16.43 & 0.058 & 0.298 & 0.360 & 0.313 & 253.68 & 0.686 & 0.161\\ 
SEIY238  & 15.88 & 0.058 & 0.275 & 0.360 & 0.298 & 237.52 & 0.686 & 0.161\\ 
SEIY220  & 15.26 & 0.055 & 0.250 & 0.342 & 0.298 & 219.87 & 0.686 & 0.161
\label{tabcritical}
\end{tabular}
\end{ruledtabular}
\end{table}

\begin{figure}[!htb]
\centering
\includegraphics[scale=0.34]{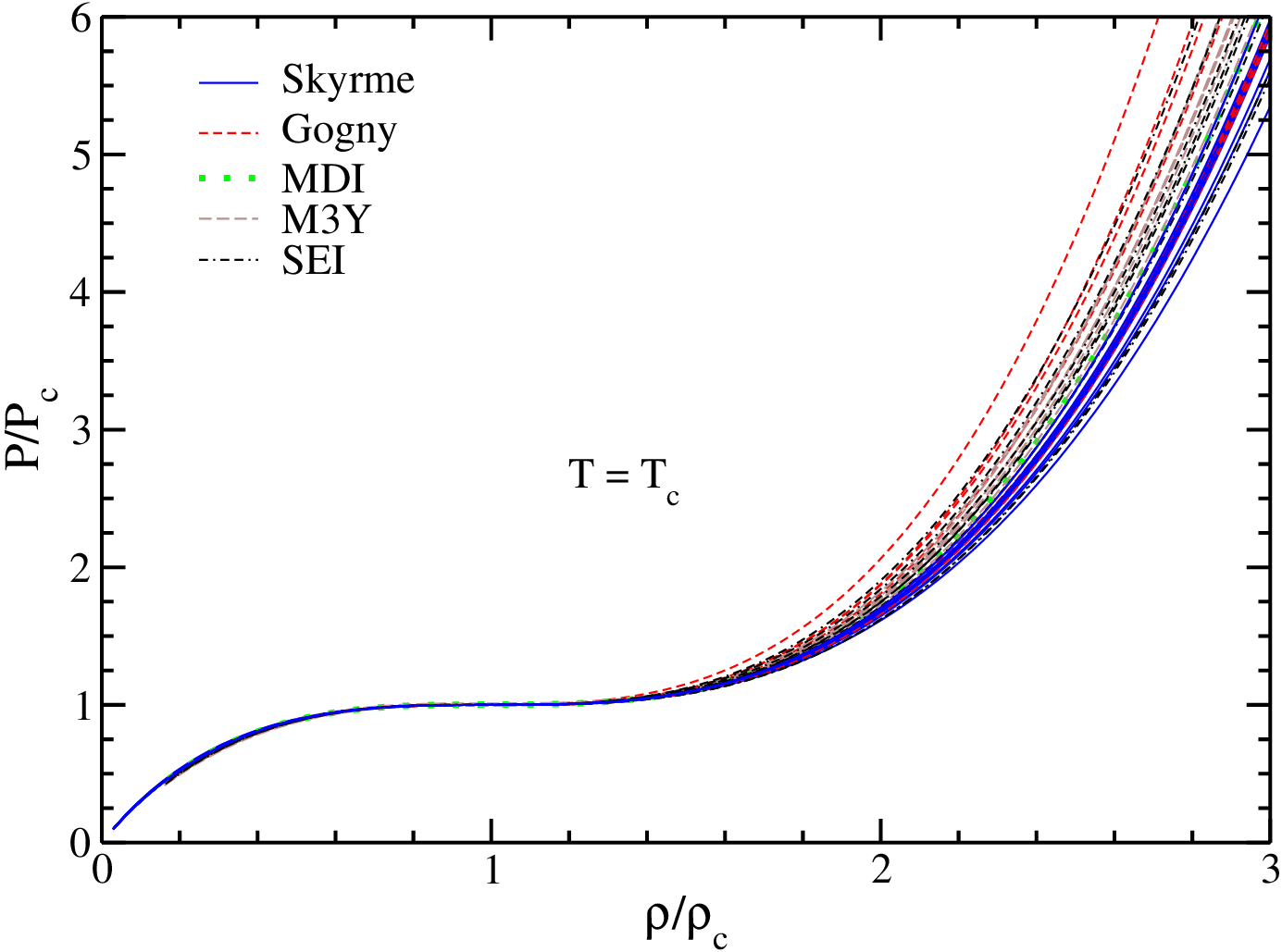}
\vspace{-0.2cm}
\caption{$P/P_c$ as a function of $\rho/\rho_c$ for the nonrelativistic parametrizations. All 
isotherms are calculated at $T=T_c$.} 
\label{prho}
\end{figure}

In Fig. \ref{prho} we present the critical isotherms i.e. pressure as a function of density 
scaled by their critical values at $T=T_c$ for different type of parametrizations considered in 
this 
work.  
It is clear that such scaled curves are indistinguishable in the gaseous phase 
($\rho<\rho_c$), and model dependent for the liquid phase ($\rho>\rho_c$), where the interactions 
become more important due to the closer proximity between the nucleons. Previously, this finding 
was observed only for those parametrizations of the relativistic mean field models which 
contain self-interactions in the scalar field $\sigma$ (Boguta-Bodmer model)~\cite{vdw5}. Later on, 
in Ref.~\cite{PRC95-065212}, it was investigated in a more sophisticated version of the RMF model 
including quartic self-interaction in the vector field $\omega_\mu$, interactions between scalar 
and vector fields ($\sigma$ and $\omega_\mu$), and interactions between scalar and isovector fields 
($\sigma$ and $\vec{\rho}_\mu$). The same pattern was observed also for those parametrizations. 
Here we observe similar findings once again for the nonrelativistic models. This  
strongly suggests towards a universality in the isotherms of symmetric nuclear matter for hadronic 
models , i.e., model independence in the gaseous region and distinguishability among the 
different interactions in the liquid phase. 

\begin{figure}[!htb]
\centering
\includegraphics[scale=0.34]{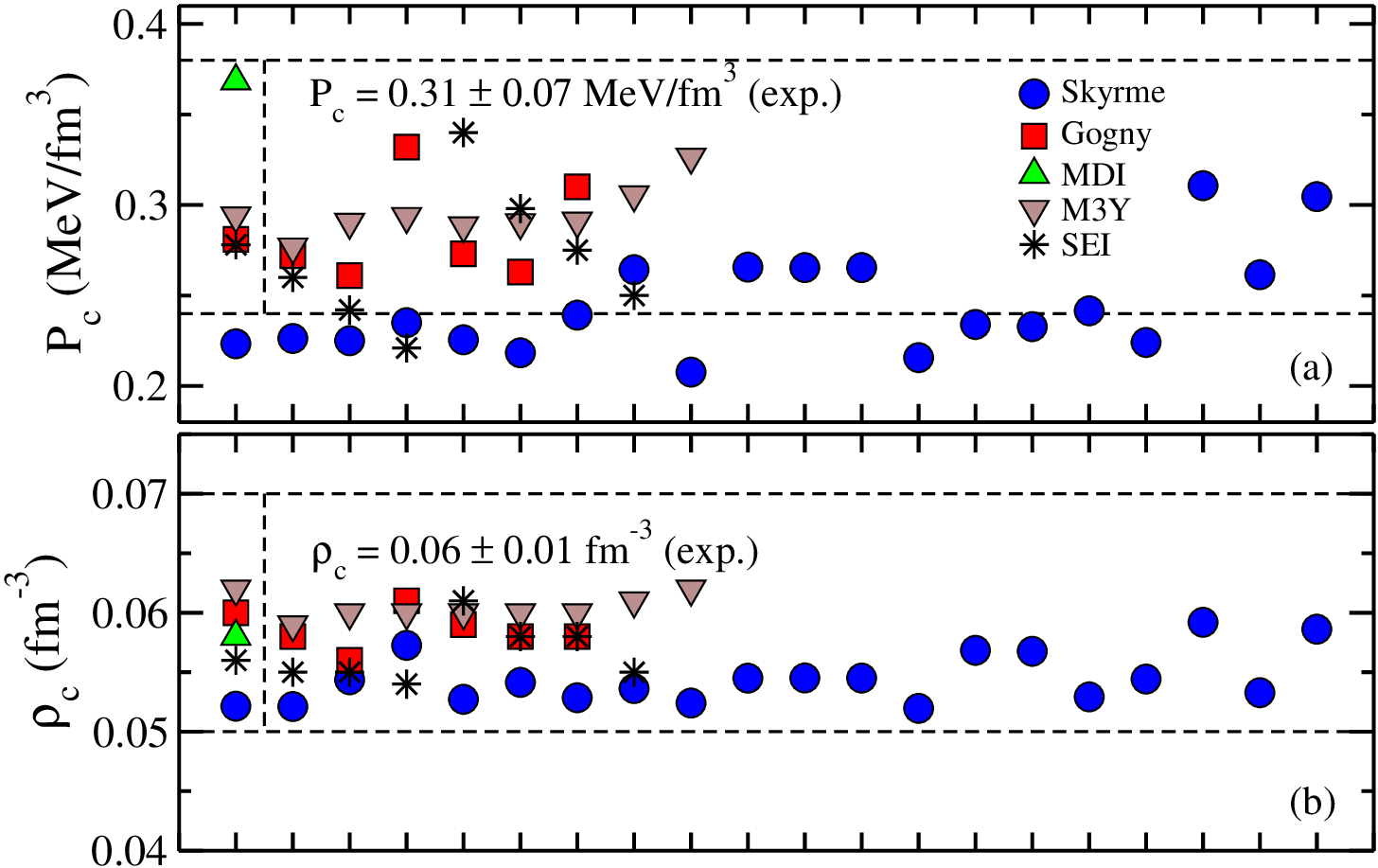}
\vspace{-0.2cm}
\caption{Values of (a) $P_c$ and (b) $\rho_c$ for the parametrizations of the Skyrme, Gogny, MDI, 
M3Y and SEI models in comparison with the corresponding experimental values extracted from 
Ref.~\cite{elliott} (ranges limited by the horizontal dashed lines).} 
\label{pc-rhoc}
\end{figure}

Concerning the critical parameters calculated for the different nonrelativistic parametrizations 
explored here, we compare our results with experimental and theoretical predictions available in 
the literature. An experimental study given in Ref.~\cite{elliott} provides values for all three 
quantities, namely,  $T_c=(17.9\pm0.4)$~MeV, $P_c=(0.31\pm0.07)$~MeV/fm$^3$, and 
$\rho_c=(0.06\pm0.01)$~fm$^{-3}$. For this purpose, the authors analyzed data from compound-nucleus 
and nuclear multifragmentation \cite{lbnl,bnl}. In Fig.~\ref{pc-rhoc} we display the outcomes 
related to $P_c$ and $\rho_c$ obtained for all nonrelativistic parametrizations used in this work.
As one can see, all the finite-range parametrizations of the Gogny, MDI, M3Y and SEI type are in 
full agreement with the experimental ranges of Ref.~\cite{elliott} for $P_c$ and $\rho_c$. On 
the other hand, the Skyrme parametrizations are also inside the range of $\rho_c$ but not all of 
them are compatible with the $P_c$ values. Ten out of twenty, namely, GSkI, GSkII, KDE0v1, MSL0, 
NRAPR, SKRA, Skxs20, SQMC650, SQMC700, and Sk$\Lambda 267$ {lie} below the lower experimental limit 
for this quantity. {It is important to mention that the effective mass seems to play an 
important role in this case. Notice that with exception of Skxs20, all the remaining 
parametrizations mentioned just above present $m^*\leqslant 0.80$}. 

\begin{figure}[!htb]
\centering
\includegraphics[scale=0.34]{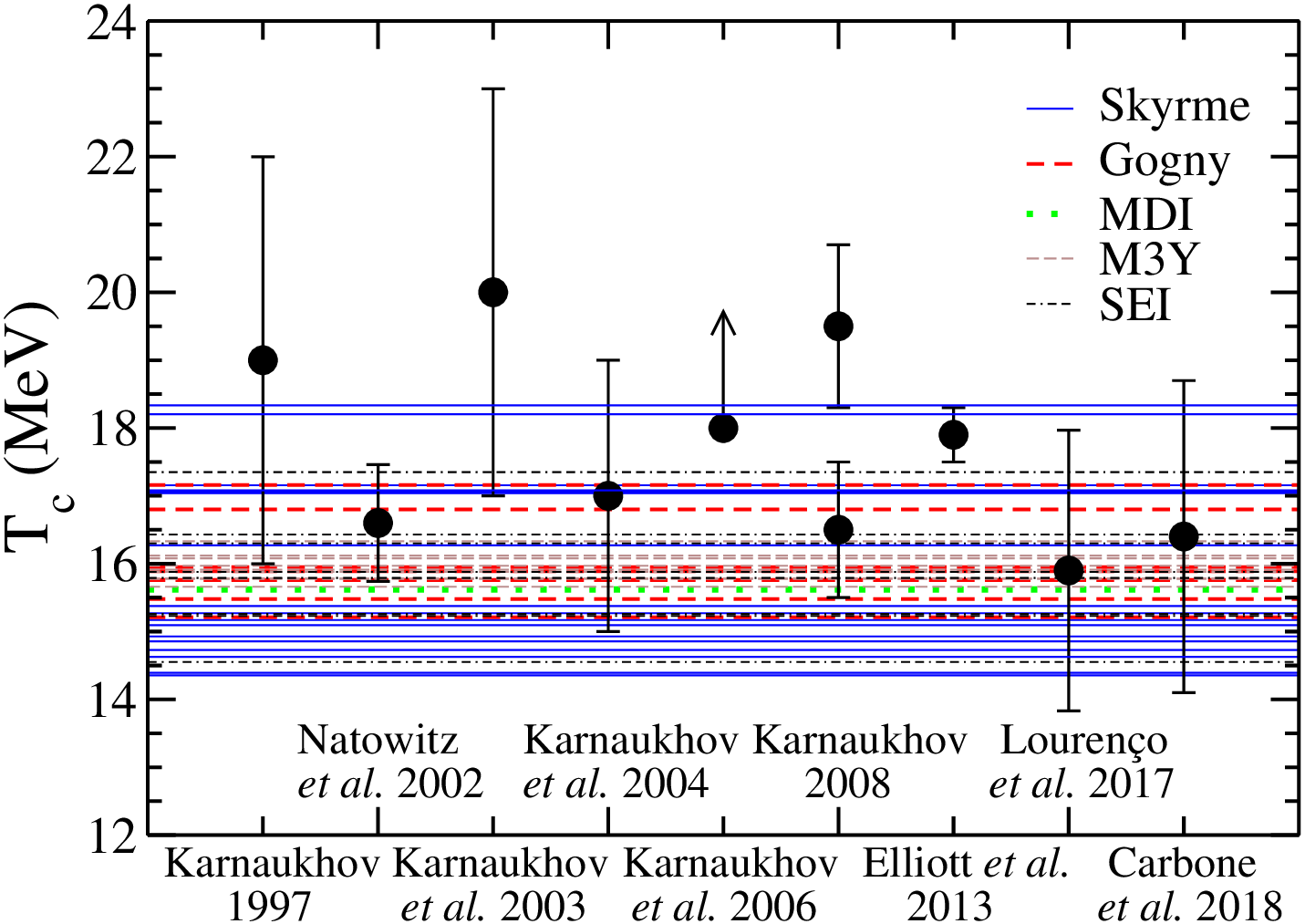}
\vspace{-0.2cm}
\caption{Values of $T_c$ for the nonrelativistic parametrizations compared with 
experimental/theoretical data (circles) collected from: Karnaukhov 1997~\cite{karn1}, Natowitz {\it 
et al.} 2002~\cite{natowitz}, Karnaukhov {\it et al.} 2003~\cite{karn2}, Karnaukhov {\it et al.} 
2004~\cite{karn3}, Karnaukhov {\it et al.} 2006~\cite{karn4}, Karnaukhov 2008~\cite{karn5}, Elliott 
{\it et al.} 2013~\cite{elliott}, Louren\c co {\it et al.} 2017~\cite{PRC95-065212}, and Carbone 
{\it et al.} 2018~\cite{PRC98-025804}.} 
\label{tc}
\end{figure}

In Fig.~\ref{tc} we display the values of $T_c$ calculated from the models analyzed in this work 
along with their different experimental values for comparison.
As one can see from this figure, the nonrelativistic parametrizations predict $T_c$ compatible 
with experimental values of Refs.~\cite{karn1,natowitz,karn2,karn3,karn4,karn5,elliott}. 
Furthermore, we also observe agreement between the results obtained with different non-relativistic 
parametrizations and the ones obtained with different theoretical models 
in Refs.~\cite{PRC95-065212,PRC98-025804}. Actually, the critical temperatures 
provided by the non-relativistic interactions analyzed in this work agree better with experimental 
values than in those of { some} RMF models (see figure 2 of Ref.~\cite{PRC95-065212}). In 
Ref.~\cite{PRC95-065212}, authors calculate $T_c$ for a class of RMF models~\cite{rmfligo} 
containing nonlinear $\sigma$ and $\omega_\mu$ terms and crossing terms involving these fields (30 
parametrizations), and for RMF models in which couplings are density dependent (4 parametrizations), 
all of them are consistent with nuclear matter constraints. In Ref.~\cite{PRC98-025804}, calculations 
were performed by using two- and three-body nuclear interactions consistently derived through chiral 
effective field theory. A van der Waals pattern was also observed in such 
models~\cite{PRC98-025804}.

\begin{figure}[!htb]
\centering
\includegraphics[scale=0.32]{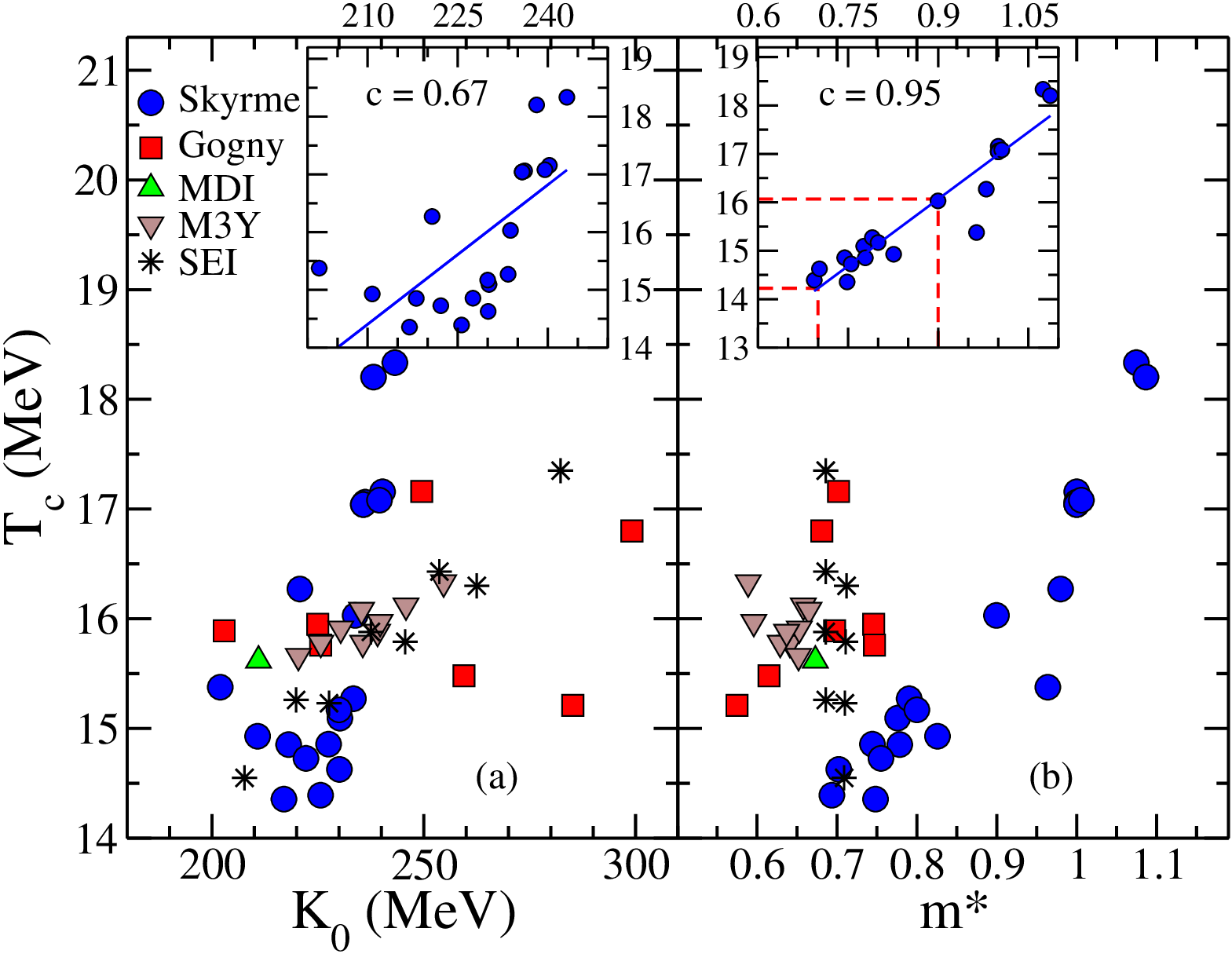}
\vspace{-0.2cm}
\caption{$T_c$ as a function of (a) $K_0$ and (b)~$m^*$ for the nonrelativistic models used in 
this work. Full lines in the insets: fitting curves related to the Skyrme parametrizations, with 
correlation coefficients given by $c=0.67$ and $c=0.95$, respectively for panels (a) and (b) (see text for 
more details).} 
\label{corr-tc}
\end{figure}

Another interesting investigation on the warm nuclear matter is the search for possible 
correlations between bulk parameters of SNM, evaluated at $\rho=\rho_0$, and the critical 
parameters. This 
feature can be useful in order to consolidate the constraints on $T_c$, $P_c$ and $\rho_c$. In 
Ref.~\cite{PRC95-065212}, for instance, it was shown that the consistent RMF models {exhibit} 
a general trend of correlation between the critical parameters and the incompressibility coefficient $K_0$. 
For the nonrelativistic parametrizations used {here, we present $T_c$ as a 
function of $K_0$ and $m^*$ in Fig.~\ref{corr-tc}.}
One can notice an increasing trend of $T_c$ as a function of both, $K_0$ and $m^*$, for the Skyrme 
parametrizations. The correlation coefficients are $0.67$ and $0.95$, for $K_0$ and $m^*$, 
respectively. The {results} observed for the Skyrme models are in line with other studies performed 
with different hadronic models. For instance, in Ref.~\cite{vdw6} the same correlation of 
Fig.~\ref{corr-tc}{\color{blue}a} is found for a class of real gas models used to describe 
symmetric nuclear matter at finite temperature, after a suitable conversion of these classical 
models into quantum ones through the incorporation of the Fermi-Dirac distribution function in the 
momentum integrals. Furthermore, we also observe qualitative agreement with other theoretical 
calculations that provide analytical expressions of $T_c$ as a function of $K_0$, as in 
Refs.~\cite{natowitz,kapusta,swesty,rios}. With regard to the $T_c$ as a function of $m^*$, we 
remark that a systematic study was performed with parametrizations of the RMF model with third- and 
fourth-order self-interactions in the scalar field $\sigma$~\cite{rmft}. For these models, we remind 
the reader that $m^*$ is the Dirac effective mass, which is slightly different from the quantity 
defined in nonrelativistic approach used in the present paper. In Ref.~\cite{rmft}, in which it was also observed that $T_c$ 
depends on $K_0$, authors verified a clear relationship between $T_c$ and $m^*$. However, {in these 
models} only the variation of $m^*$ was taken into account, {\it i.e.}, saturation density, binding 
energy and incompressibility were kept fixed. Regarding the finite-range models, we can see that  
the SEI family of parametrizations, which  have very similar nuclear matter properties except 
incompressibility, show a very clear correlation between $T_c$  and $K_0$ with a correlation 
coefficient of 0.98 (see Table~\ref{corrtab}). This correlation is  also observed in the M3Y 
parametrizations where a correlation coefficient of 0.89 was found. Concerning the relation between 
the $T_c$ and $m^*$, the predictions of the finite-range interactions, in particular SEI and M3Y, 
and the ones of the Skyrme forces are clearly different. The SEI interactions, and to some extent 
the M3Y ones, have almost the same effective mass and there is no correlation between $T_c$ and 
$m^*$. This situation is different from the one found with the Skyrme forces, where a clear linear 
correlation is observed (see the inset of Figure \ref{corr-tc}{\color{blue}b}). However, for Gogny 
forces, which have properties in symmetric nuclear matter quite different {among them}, do not show any clear 
correlation between  $T_c$ and $K_0$ or $m^*$ (see also Ref. \cite{rios} in this respect).

One needs to be careful in a correlation study like the present one. Some of the finite range 
interactions used in the present work were obtained in a systematic way to satisfy certain 
constraints. Their merits should not be tested only with a correlation study. However, 
{most of the Skyrme parametrizations} used in the present work satisfy several 
independent constraints imposed by experiments and {astronomical} observations (see 
Ref.~\cite{PRC85-035201}). An independent correlation study is quite justified using only these 
Skyrme interactions. In the inset of Fig.~\ref{corr-tc}{\color{blue}b} one can  
see a positive linear correlation between $T_c$ and $m^*$. If we take a conservative 
estimate of $m^*$ of $0.7$ - $0.9$, it translates into {a variation of} $T_c$ from $14.225$~MeV to $16.066$~MeV, as 
$T_c$ and $m^*$ show a high positive correlation between them. These are indicated by the red 
parallel lines to the axes in the inset of Fig.~\ref{corr-tc}{\color{blue}b}.
\begin{table}[!htb]
\begin{ruledtabular}
\caption{Correlation coefficients ($c$) among different pairs of critical parameters and nuclear 
matter properties are listed for four families of non-relativistic interactions considered in this 
work along with combining them together in ``all".}
\centering
\begin{tabular}{c|r|r|r|r|r}
$c$                & Skyrme  & Gogny  & M3Y    & SEI    & all  \\
\hline
$T_c\times K_0$    & 0.67    &  0.13  &  0.89  &  0.98  &  0.44 \\ 
$T_c\times m^*$    & 0.95    &  0.45  & -0.37  & -0.43  &  0.51 \\ 
$P_c\times K_0$    & 0.68    &  0.26  &  0.78  &  0.93  &  0.58 \\
$P_c\times m^*$    & 0.83    &  0.16  & -0.81  & -0.56  & -0.20 \\
$\rho_c\times K_0$ & 0.17    & -0.11  &  0.42  &  0.80  &  0.33 \\ 
$\rho_c\times m^*$ & 0.26    & -0.16  & -0.64  & -0.67  &  0.49 \\
$T_c\times P_c$    & 0.95    &  0.94  &  0.79  &  0.98  &  0.72
\label{corrtab}
\end{tabular}
\end{ruledtabular}
\end{table}

\begin{figure}[!htb]
\centering
\includegraphics[scale=0.32]{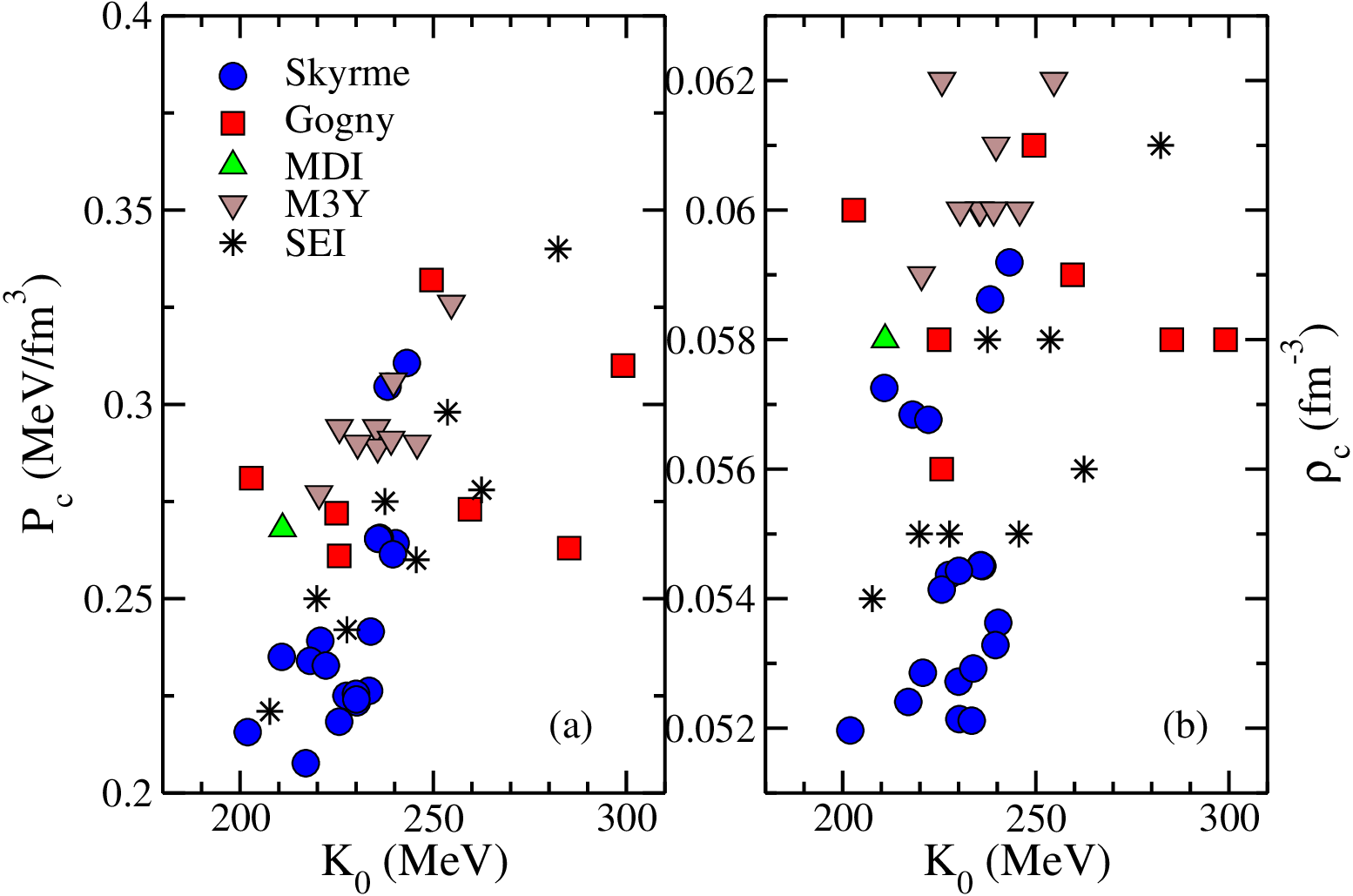}
\vspace{-0.2cm}
\caption{Critical (a) pressure and (b) density as a function of~$K_0$ for the different Skyrme, 
Gogny, MDI, M3Y and SEI parametrizations considered in this work (see text for details).}
\label{corr-pc-rhoc}
\end{figure}

The influence of $K_0$ on $P_c$ and $\rho_c$ is analyzed in Fig.~\ref{corr-pc-rhoc}. 
The {behavior} of $P_c$ and $\rho_c$ as increasing functions of $K_0$ was {also} observed 
for the RMF models investigated in Refs.~\cite{PRC95-065212,rmft}. From the Fig.~\ref{corr-pc-rhoc} 
we can still appreciate the correlations between $P_c$ and $K_0$, in particular for the SEI and M3Y 
forces and less  clearly  for the Skyrme interactions. This  is confirmed by the correlation 
coefficients reported  in Table \ref{corrtab}. From Fig. \ref{corr-pc-rhoc}{\color{blue}a} and Table 
\ref{corrtab} it is again clear that Gogny forces do not show $P_c$ - $K_0$ correlation. The 
results reported in Table \ref{corrtab} also show that there is no correlation between $K_0$ or 
$m^*$ and $\rho_c$.
\begin{figure}[!htb]
\centering
\includegraphics[scale=0.31]{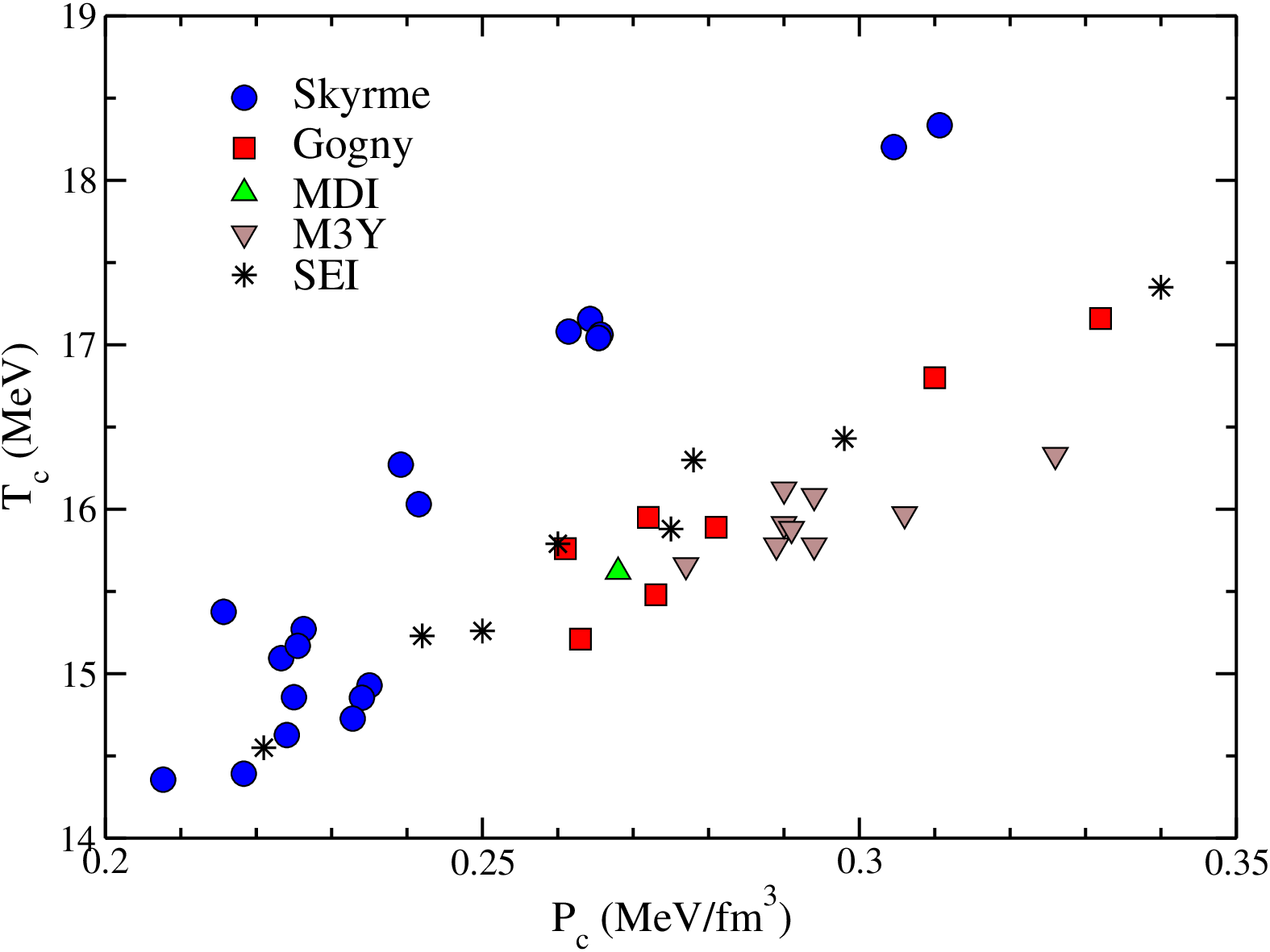}
\vspace{-0.2cm}
\caption{$T_c$ as a function of $P_c$ for the different Skyrme, Gogny, MDI, M3Y and SEI 
parametrizations considered in this work (see text for more details).} 
\label{corr-tc-pc}
\end{figure}

Finally, we display in Fig.~\ref{corr-tc-pc} the relationship between $T_c$ and $P_c$. For the 
classical van der Waals model, one has $T_c=8bP_c$ with $b$ being the excluded volume parameter 
(strength of the repulsive interaction), indicating a clear linear relation. For the 
nonrelativistic models studied here, an increment of $T_c$ as a function of $P_c$ is observed with 
some deviation from the exact linear pattern. A much more clear linear behavior was observed, for 
instance, with the RMF parametrizations and density-dependent RMF Hartree-Fock models 
used in Ref.~\cite{yang}. All the finite-range models follow the $T_c$ - $P_c$ correlation quite precisely 
as it can be seen from Fig.~\ref{corr-tc-pc} and from the correlation coefficients given in 
Table~\ref{corrtab}. Regarding the Skyrme results, we see that the models which predict a critical 
pressure below $0.24$~MeV.fm$^{-3}$ and at the same time have a small effective mass $m^*$ below 
0.8 are well aligned with the finite-range ones. We also see that the remaining Skyrme 
parametrizations, which have an effective mass close to the {bare mass}, lie on top with an other 
parallel line shifted to higher critical temperature. As a consequence, our study predicts that the 
$T_c$ - $P_c$ correlation is reinforced for models with similar effective mass. One can notice that, the Skyrme 
models just mentioned above with $m^*$ greater than $0.8$, reproduce better the experimental 
constraint on the critical pressure $P_c$ (see Fig.~\ref{pc-rhoc} and corresponding discussion). 
However, they follow a different $T_c$-$P_c$ correlation line compared to the rest of models 
considered in this work including the ones of the Skyrme family (see Fig.~\ref{corr-tc-pc}).


\section{Summary and conclusions}
\label{summary}

In this work we have analyzed symmetric nuclear matter at finite temperature for a set of 
parametrizations of the Skyrme, Gogny, MDI, M3Y and SEI nonrelativistic models. For the first one, 
we have chosen the so called consistent Skyrme parametrizations (CSkP), namely, GSkI, GSkII, 
KDE0v1, LNS, MSL0, NRAPR, Ska25s20, Ska35s20, SKRA, Skxs20, SQMC650, SQMC700, SkT1, SkT2, SkT3 and 
SV-sym32. They satisfy a set of constraints related to the nuclear matter and pure 
neutron matter \cite{PRC85-035201}. Furthermore, they are also consistent with the boundaries of 
the tidal deformabilities determined by the LIGO and Virgo Collaboration studies, all of them 
related to the detection of gravitational waves coming from the neutron star merger event 
GW170817~\cite{epja}. For the finite-range models, we have chosen some representative 
parametrizations. We also furnished the expressions for the pressure as a function of temperature 
and density for the {considered} models (see Eqs.~(\ref{presskyrme}) and~(\ref{press})). Once this 
thermodynamical quantity was determined, it was possible to find the critical parameters (CP) of the models, 
namely, $T_c$, $P_c$ and $\rho_c$, by imposing the conditions given in Eq.~(\ref{conditions}). The 
respective values of these quantities are listed in Table~\ref{tabcritical}. To the best of our 
knowledge, assembly of critical properties of nuclear matter at finite temperature of this sort are 
quite scarce in the literature for finite-range interactions.

One of the results found in our investigation is the pattern exhibited in Fig.~\ref{prho}, namely, 
all isotherms collapse in the low density region (gaseous phase), and nuclear interactions 
become important for densities greater than $\rho_c$ (liquid phase). Such a feature was also 
observed in Ref.~\cite{vdw5} in which the calculations were restricted to the relativistic mean 
field (RMF) model presenting third- and fourth-order self-interactions in the scalar field. 
In Ref.~\cite{PRC95-065212}, the same phenomenology was observed for more 
sophisticated version of the RMF models including quartic self-interaction in the vector field 
and other mesonic interactions. Our finding strongly suggests a kind of universality for 
the isotherms of the hadronic models (relativistic and nonrelativistic) for symmetric nuclear matter.

With regard to the values of the CP of the studied nonrelativistic parametrizations, we found very good 
agreement of all the models with the experimental value of 
$\rho_c=(0.06\pm0.01)$~fm$^{-3}$~\cite{elliott}. All the finite-range models considered here as well 
as ten out of twenty Skyrme parametrizations analyzed, lie within the experimental 
limit of $P_c=(0.31\pm0.07)$~MeV/fm$^3$~\cite{elliott}. Finally, concerning $T_c$, we {have} 
compared our {results} with both, theoretical and experimental data collected from 
the literature. {Fig.~\ref{tc} shows that all the models analyzed here} are compatible with the 
experimental values. {We have also} verified that the {critical parameters} obtained here are also 
compatible with { previous} {theoretical results}
 {reported} in Refs.~\cite{PRC95-065212,PRC98-025804}, in 
which the authors have used a class of RMF models~\cite{PRC95-065212}, and {with the} nuclear models  
derived from chiral effective field theory~\cite{PRC98-025804}.

Another investigation performed in this work was the search for {possible} correlations between 
bulk parameters, evaluated at the saturation density, and the CP.  In Table~\ref{corrtab} we 
have shown  the correlation coefficients obtained for some possible relationships. The general trend of 
$T_c$ as an increasing function of $K_0$ was found for all families of models individually with the exception of 
the Gogny parametrizations. This finding is compatible with studies using other hadronic 
models~\cite{vdw6,natowitz,kapusta,swesty,rios,rmft}. In Fig.~\ref{corr-pc-rhoc}{\color{blue}a} 
and in Table~\ref{corrtab}, where the correlation coefficient for the $P_c\times K_0$ 
relationship was presented, we observed the same pattern, namely, $P_c$ and $K_0$ 
are correlated to each other for all 
models except for the Gogny ones. This particular correlation was also exhibited for relativistic 
parametrizations explored in Refs.~\cite{PRC95-065212,rmft}). With regard to $\rho_c$ as a function 
of $K_0$, we find a good correlation coefficient only for the parametrizations of the SEI model, 
namely, $c=0.80$. Concerning the critical parameters as a function of $m^*$, we found hints of 
correlation for Skyrme~($T_c$ and $P_c$), M3Y~($P_c$ and $\rho_c$) and SEI~($\rho_c$) models.
Finally, in Fig.~\ref{corr-tc-pc} we verified correlation between $T_c$ and $P_c$ in agreement with 
other findings~\cite{yang}. Also our results seems to point out that this {specific} 
correlation is better fulfilled for models with  similar effective mass independently  of the type 
of interaction considered.

\section*{Acknowledgments} 

This work is a part of the project INCT-FNA Proc. No. 464898/2014-5, partially supported by 
Conselho Nacional de Desenvolvimento Científico e Tecnológico (CNPq) under grants 310242/2017-7, 
406958/2018-1, 312410/2020-4 (O.L.) and 433369/2018-3 (M.D.). We also acknowledge Funda\c{c}\~ao 
de Amparo \`a Pesquisa do Estado de S\~ao Paulo (FAPESP) under Thematic Project 2017/05660-0 (O.L., 
M.D.) and Grant No. 2020/05238-9 (O.L., M.D.). The work of X.V. and C.M. was partially supported by 
Grant FIS2014-54672-P from MINECO and FEDER, Grant 2014SGR-401 from Generalitat de Catalunya, and 
from the State Agency for Research of the Spanish Ministry of Science and Innovation through the 
Unit of Excellence Maria de Maeztu 2020-2023 award to the ICCUB (CEX2019-000918-M). C.M. also 
acknowledges the financial support from CEFIPRA project number 5804-3. O.L. and M.D. 
also thank A. S. Schneider for the very fruitful discussions regarding the Skyrme model at finite 
temperature.

\section{Appendix}

In momentum space, the finite-range interaction is given by the Fourier transform of the form 
factor in coordinate space $f(s,\mu)$, where $s=\vert {\bf r} - {\bf r'}\vert$ and $\mu$ is the 
range of the force, and, therefore
\begin{eqnarray}
g(\vert {\bf k} - {\bf k'}\vert,\mu) = \int d{\bf s} e^{i({\bf k} - {\bf k'}){\bf s}} f(s,\mu).
\end{eqnarray}
The interaction in momentum space depends on the modulus of the relative momentum and therefore on 
the angle between {\bf k} and {\bf k$^\prime$}. We can finally write the interaction with spherical 
symmetry in momentum space by performing the angular average: 
\begin{eqnarray}
\tilde{g}(k,k',\mu) = \frac{1}{4\pi}\int d\Omega g(\sqrt{k^2+{k'}^2-2kk'\cos{\theta}}). 
\end{eqnarray} 
In this work we use Gaussian $f_G(s)=e^{-s^2/\mu^2}$ and Yukawian $f_Y(s)=e^{-\mu s}/\mu s$ 
form-factors. The corresponding Fourier transforms are
\begin{eqnarray}
g_{\tiny G}(\vert {\bf k} - {\bf k'}\vert,\mu) = (\sqrt{\pi}\mu)^3 e^{-\frac{\mu^2({\bf k} 
- {\bf k'})^2}{4}}
\end{eqnarray}
and 
\begin{eqnarray}
g_{\tiny Y}(\vert {\bf k} - {\bf k'}\vert,\mu) = \frac{4\pi}{\mu} \frac{1}{\mu^2 + ({\bf k} 
- {\bf k'})^2}.
\end{eqnarray}
After the angular average they become
\begin{eqnarray}
\tilde{g}_{\tiny G}(k,k',\mu) = \frac{2\pi^{3/2}\mu}{k k'}e^{-\frac{\mu^2(k^2 + 
{k'}^2)}{4}}\sinh{\frac{\mu k k'}{2}}
\end{eqnarray}
and
\begin{eqnarray}
\tilde{g}_{\tiny Y}(k,k',\mu) = \frac{\pi}{\mu k k'}\ln{\frac{\mu^2 + (k + k')^2}{\mu^2 
+ (k-k')^2}},
\end{eqnarray}
which enter in Eqs.~(\ref{spe}) and~(\ref{ener}).

\end{document}